\documentclass{jetpl}
\usepackage{color,amsmath,amsfonts,graphics,epsfig,amssymb}
\usepackage[english]{babel}
\usepackage{xcolor}
\definecolor{xlinkcolor}{cmyk}{1,1,0,0}
\usepackage[bookmarks=true, pdfnewwindow=true, colorlinks=true, linkcolor=xlinkcolor, citecolor=xlinkcolor, filecolor=xlinkcolor, urlcolor=xlinkcolor, final=true]{hyperref}
\twocolumn
\lat
\newcommand{\red}{\color{black}}
\newcommand{\black}{\color{black}}
\title{
Testing new-physics scenarios with the combined LHAASO and Carpet-3 fluence spectrum of GRB~221009A:\\
axion-like particles and Lorentz-invariance violation
}

\rtitle{New physics and GRB~221009A}

\sodtitle{Testing new-physics scenarios with the combined LHAASO and Carpet-3 fluence spectrum of GRB~221009A: axion-like particles and Lorentz-invariance violation}

\author{P.\,S.\,Satunin$^{a}$
and S.\,V.\,Troitsky$^{a,b,}$\thanks{E-mails: satunin@inr.ac.ru, st@inr.ac.ru}}

\rauthor{P.\,S.\,Satunin, S.\,V.\,Troitsky}

\sodauthor{P.\,S.\,Satunin, S.\,V.\,Troitsky}

\address{$^{a}$Institute for Nuclear Research of the Russian Academy of
Sciences,\\
60th October Anniversary prospect 7A, 117312 Moscow, Russia\\
and\\
$^{b}$Faculty of Physics, Lomonosov Moscow State University, 1-2 Leninskiye Gory, 119991 Moscow, Russia}

\dates{October 10, 2025}{November 5, 2025}

\abstract{
From gamma-ray burst (GRB) 221009A, very high-energy photons were detected: $\gtrsim 10$ TeV with LHAASO and $\gtrsim 100$ TeV with Carpet-3. Such energetic photons are expected to be absorbed via electron-positron pair production on their way to the Earth. Their observation might be explained by new physics, including Lorentz invariance violation (LIV) or photon mixing with axion-like particles (ALPs). Here, we construct a joint fluence spectrum by combining flux measurements from both experiments, and fit it under these hypotheses. \red While \black LIV can account for the Carpet-3 observation, \red it provides only \black a modest improvement over standard physics in the overall fit \red and requires parameters excluded by other constraints\black. ALP mixing improves the description of both LHAASO and Carpet-3 data, yielding a substantial enhancement in fit quality for a specific region of the ALP parameter space.
}
\begin{document}
\maketitle
\sloppy
\noindent\textbf{1. Motivation.} 
The brightest gamma-ray burst (GRB) ever detected, GRB~221009A, immediately attracted the attention of the particle-physics community due to observation of extremely energetic photons from this distant explosion. Shortly after the burst, the LHAASO experiment reported \cite{LHAASO-GCN} an association of photons with energies up to 18~TeV with the event, while Carpet-2 claimed \cite{CarpetATel-GRB} the observation, roughly one hour later, of a single photon-like event with an energy of about 250~TeV. The association of both 18- and 250-TeV photons with a source at the burst redshift, $z = 0.151$, appeared problematic, as such photons should have been absorbed via $e^+e^-$ pair production on the cosmic background radiation \cite{Nikishov}.

Possible explanations invoking new physics that could modify gamma-ray propagation were proposed almost immediately. Among them were photon mixing with axion-like particles (ALPs) \cite{Galanti:2022,ST-GRB-JETPL} and Lorentz invariance violation (LIV) \cite{Finke:2022LIV}, both of which were consistent with the limited observational information available at that time. Subsequently, both collaborations published detailed accounts of their observations in refereed journals \cite{LHAASO-WCDA-GRB,LHAASO-KM2A-GRB,CarpetGRB-PRD}. The recent update \cite{CarpetGRB-PRD} renewed interest in new-physics interpretations \cite{Galanti:2025LIVandALP,Galanti:2025LIV,Ofengeim:2025LIV,Song:2025LIV}, yet a quantitative comparison of model predictions with the combined spectra from both experiments has remained absent.

Here, we address this gap \red and perform a quantitative \red test -- missing in previous works  -- of hypotheses of standard physics, photon–ALP mixing, and LIV. To this end, we construct \black an estimated fluence spectrum that combines the data obtained at different time intervals\red, and perform a likelihood fit of \black this spectrum under the assumptions \red of these three hypotheses. \black

\vskip 1mm\noindent\textbf{2. Scenarios.} 
\vskip 0.5mm
\noindent{\sl 2.1. Standard physics.}
Energetic gamma rays propagating through the Universe are attenuated by background radiation via the $\gamma\gamma \to e^+e^-$ process \cite{Nikishov}, provided that the combined energy of the two photons exceeds the threshold for electron–positron pair production. Throughout this work, we adopt the model\footnote{Numerical tables from Ref.~\cite{Saldana-Lopez:2020} are available at \url{https://www.ucm.es/blazars/ebl}. To reproduce the optical depths in \texttt{tau\_saldana-lopez21.out}, we assumed that the photon densities in \texttt{ebl\_saldana21\_comoving.txt} are given in the observer’s frame.} of the extragalactic optical and infrared background light (EBL) by Saldana-Lopez \textit{et~al.}\ (2021) \cite{Saldana-Lopez:2020}. We also include the contribution of the cosmic microwave background (CMB) radiation. While the CMB component—dominant for photon absorption in the Carpet-3 energy range—is well established, multiple EBL models exist. The model of Ref.~\cite{Saldana-Lopez:2020} was chosen for consistency with the analyses performed in previous LHAASO \cite{LHAASO-WCDA-GRB,LHAASO-KM2A-GRB} and Carpet-3 \cite{CarpetGRB-PRD} studies. 

\vskip 0.5mm
\noindent{\sl 2.2. ALP mixing.}
Photon propagation is modified in the presence of ALPs due to photon–ALP mixing in cosmic magnetic fields \cite{RaffeltStodolsky}; see, e.g., Refs.~\cite{ST-mini-rev,Roncadelli-review2022} for reviews and additional references. For the range of ALP parameters where strong mixing is expected in both the LHAASO and Carpet-3 energy bands \cite{ST-GRB-JETPL}, the effect of the weak \cite{Pshirkov-IGMF} intergalactic magnetic fields can be safely neglected, and only mixing within the GRB host galaxy and the Milky Way needs to be considered. We adopt the Galactic magnetic-field model of Ref.~\cite{Pshirkov-GMF} and the host-galaxy model of Ref.~\cite{ST-GRB-hostMF}. The latter includes two parameters referred to as \textit{unknown unknowns}; we follow the procedure of Ref.~\cite{ST-GRB-hostMF} to compute a weighted average over them.

Photon–ALP mixing is treated using the density-matrix formalism described e.g.\ in Ref.~\cite{LT-magnetic-fields}, while photon attenuation is accounted for as described in Sec.~2.1.

\vskip 0.5mm
\noindent{\sl 2.3. LIV.}
Presence of tiny effects of LIV may be parametrized phenomenologically with a modified dispersion relation for photons,
\begin{equation}
\label{eq:MDR}
    E^2 - \mathbf{k}^2 = s_n\frac{E^{n+2}}{E_{\text{LIV},n}^n},
\end{equation}
where $E$ and $\mathbf{k}$ are the energy and the 3-momentum of the photon, $n=1 (2)$ relates to the linear (quadratic) LIV, respectively. The energy scale of LIV is denoted by $E_{\text{LIV},n}$, while the sign $s_n = +1(-1)$ relates to the cases of super(sub)luminal LIV. In what follows, we concentrate on the case of $n=2$ subluminal LIV; other options are strongly constrained experimentally \cite{Addazi:2021xuf}. 

In the presence of LIV, cross sections of particle processes, including the one responsible for the attenuation, $\gamma\gamma \to e^+e^-$, get modified. In the literature, these modifications were approximated by different ways, including only the modification of the threshold \cite{Blanch:2001hu, Jacob:2008gj} or a shift in the Mandelstam variable $s$ in the cross section \cite{Fairbairn:2014LIV}, see e.g.\ \cite{Tavecchio:LIV-compare} for a discussion. Recently, the $n=2$ LIV cross section has been calculated explicitly % in EFT framework 
\cite{thesisCrossSec}. 

The standard Lorentz-invariant Mandelstam variable is defined as  
$$
s = 4m_e^2 s_0 \equiv (k_1^\mu + k_2^\mu)^2 = 2 E_1 E_2 (1 - \cos\theta),
$$
where $k_{1,2}^\mu$ are the four-momenta of the two photons, and $\theta$ is the angle between their three-momenta.  
In the case of LIV, the modified $s$ becomes  
$$
(k_1^\mu + k_2^\mu)^2 \equiv 4m_e^2 s_1 = 4m_e^2 (s_0 - \Delta),
$$
where $\Delta = E^4 / (4m_e^2 E_{\rm LIV,2}^2)$.  
The corresponding cross section~\cite{thesisCrossSec} is given by  
\begin{align}
\label{eq:exactCS}
&\sigma_\text{LIV}\!= \!\frac{ \pi \alpha^2}{2m_e^2 s_0} \Biggl[\left(2+\frac{2s_1(1-2\Delta)}{s_0^2}-\frac{(1-\Delta)}{s_0^2}\right)\! \times \notag \\ & 
\!\times \!\ln\!\left(\frac{1\!+\!\sqrt{1-1/s_1}}{1\!-\!\sqrt{1-1/s_1}}\right)\!
- \!\left(2\!+\!\frac{2s_1(1-4\Delta)}{s_0^2}\right) \!\!\sqrt{1-\frac{1}{s_1}} \Biggr].\!\!
%\label{eq:F_complete}
\end{align}
The optical depth $\tau$ corresponding to this cross section was analyzed in Ref.~\cite{Carmona:2024correctCrossSec}.  
In the present work, we compute $\tau$ using Eq.~\eqref{eq:exactCS}, together with the EBL and CMB photon densities described in Sec.~2.1.

\vskip 1mm\noindent\textbf{3. Fluence spectrum construction.} 

\vskip 0.5mm
\noindent{\sl 3.1. LHAASO.}
The experiment comprises two largely independent detector systems: the Water Cherenkov Detector Array (WCDA), sensitive to photon energies of approximately 0.1$-$10~TeV, and the Kilometer Square Array (KM2A), operating in the 10$-$1000~TeV range. GRB~221009A was observed by both arrays, and the corresponding fluxes have been published in Refs.~\cite{LHAASO-WCDA-GRB,LHAASO-KM2A-GRB}.

The energy of a primary particle initiating a shower detected by a surface array is determined indirectly, and its best-fit value depends on the assumed spectral shape, even for a single event, within the energy reconstruction uncertainties. In particular, for the highest-energy photon associated with GRB~221009A, Ref.~\cite{LHAASO-KM2A-GRB} reports estimates ranging from 12.2~TeV -- assuming a power-law spectrum with an exponential cutoff -- to 17.8~TeV, as initially quoted in Ref.~\cite{LHAASO-GCN}, under the log-parabola assumption. In this work, we adopt the former, more conservative, spectral points.

\begin{figure}
    \centering
    \includegraphics[width=\linewidth]{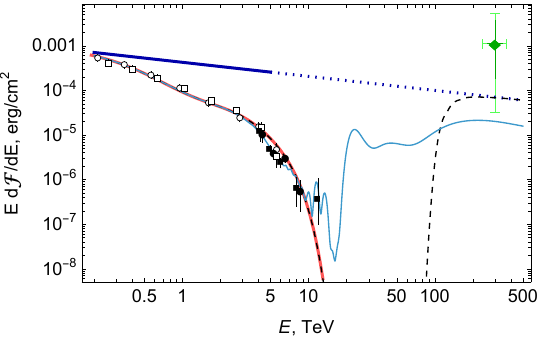}
    \caption{\label{fig:spec-fit} \sl
\textbf{Figure~\ref{fig:spec-fit}.}
Combined fluence spectrum for the entire duration of GRB~221009A. Black symbols -- LHAASO (circles: derived from 230$-$300~s, boxes: derived from 300$-$900~s; empty symbols for WCDA, filled symbols for KM2A). Green diamond -- Carpet-3 (dark error bars are 68\% CL, light error bars are 95\% CL). Dark blue line -- WCDA best-fit power-law intrinsic spectrum, extrapolated as the dotted line. Red line -- this best-fit spectrum after absorption. Thin blue line -- example of the spectrum with ALPs for the best-fit ALP parameters; dashed black line -- best-fit spectrum with LIV. 
See the text for details.}
\end{figure}

For both detector systems, fluxes are reported for two time intervals, (230$-$300~s and 300$-$900~s after the GRB trigger), yielding four independent data sets corresponding to the two arrays and two observation windows. Our goal is to obtain the total fluence, that is, the time-integrated flux. At all energies, the temporal evolution of the flux is well described by a single light curve presented in Ref.~\cite{LHAASO-WCDA-GRB}, which also fits the KM2A data~\cite{LHAASO-KM2A-GRB}. We use this light curve to reconstruct the LHAASO fluence spectrum over the full observation interval, 0$-$2000~s, combining the four individual spectra. The resulting fluence spectra are mutually consistent, as illustrated in Fig.~\ref{fig:spec-fit}.

\vskip 0.5mm
\noindent{\sl 3.2. Carpet-3.}
We use the data from Ref.~\cite{CarpetGRB-PRD}, which represent a significant update compared to the original telegram~\cite{CarpetATel-GRB}. The revised analysis includes data from the extended muon detector, which was already operating in commissioning mode on the day of the GRB—hence the transition from the name Carpet-2 to Carpet-3. It also incorporates dedicated Monte Carlo simulations that refine the energy reconstruction for this specific event direction, yielding an updated estimate of $300^{+43}_{-38}$~TeV. In addition, a new photon–hadron separation technique based on machine learning was employed to improve event classification.

The event was recorded 4536~s after the GRB trigger, when the source was leaving the LHAASO field of view due to Earth’s rotation. This temporal offset complicates the direct combination of the LHAASO and Carpet-3 flux spectra in a joint quantitative analysis, motivating the use of a fluence-based approach, which we adopt here. The resulting fluence estimate from Carpet-3 is taken from Ref.~\cite{CarpetGRB-PRD}.

\vskip 1mm\noindent\textbf{4. Fluence spectrum fits.} 

\vskip 0.5mm
\noindent{\sl 4.1. Likelihood construction.}
For model fitting, we construct a combined likelihood function that incorporates all fluence spectrum data points from the WCDA, KM2A, and Carpet-3 detectors.
The total likelihood is defined as 
\begin{equation}
\mathcal{L} = 
\prod_i P_{\rm G}(F_i^\text{obs}\!,\sigma_i; F_i^\text{mod}) \times 
\prod_j P_{\rm P}(n_j^\text{obs}\!,b_j; n_j^\text{mod}).
\end{equation}
Here, $i$ enumerates WCDA data points for which we assume Gaussian statistics, using the point-by-point fluences, $F_i^\text{obs}\pm \sigma_i$, and $P_{\rm G}$ is the Gaussian probability distribution function for the model-predicted fluences $F_i^\text{mod}$.
%\begin{equation}
%\mathcal{L}_{{\rm G},i}=
%\frac{1}{\sqrt{2\pi},\sigma_i}
%\exp\!\left[-\frac{(F_i^\text{obs}-F_i^\text{mod})^2}{2\sigma_i^2}\right],
%\end{equation}
For the KM2A and Carpet-3 data, where the number of detected events is small, we employ Poisson statistics, taking into account the observed, $n_j^\text{obs}$, and expected background, $b_j$, event numbers reported, for each energy bin $j$, in Refs.~\cite{LHAASO-KM2A-GRB,CarpetGRB-PRD}. By $P_{\rm P}$ we denote the Poisson probability distribution function 
%\begin{equation}
%\mathcal{L}_{{\rm P},j} =
%\frac{(n_j^\text{mod}+b_j)^{n_j^\text{obs}}}{n_j^\text{obs}!}
%\exp[-(n_j^\text{mod}+b_j)],
%\end{equation}
for the model-predicted event numbers $n_j^\text{mod}$.
Defined in this way, the likelihood remains finite even for models predicting zero signal flux in the Carpet-3 bin, owing to the nonzero background of 0.003~events. To determine the best-fit model parameters, we maximize the total likelihood $\mathcal{L}$ or, equivalently, minimize the test statistic $\chi^2 = -2 \ln \mathcal{L}$.

\vskip 0.5mm
\noindent{\sl 4.2. Intrinsic spectrum: WCDA fit.}
To test different scenarios affecting photon propagation, we must assume an intrinsic gamma-ray spectrum at the source. A power law is a natural choice: any concave spectrum would make it difficult to reproduce the Carpet-3 point even in the absence of absorption, while a convex spectrum lacks physical motivation. Moreover, the power law
\begin{equation}
E\, \frac{d \mathcal{F}}{dE} = 4.236 \times 10^{-4}\, {\rm erg\,cm}^{-2} \left(\frac{E}{\rm TeV}\right)^{-0.315},    
\label{eq:intrinsic}
\end{equation}
attenuated according to the Saldana-Lopez \textit{et al.}~\cite{Saldana-Lopez:2020} model, provides an excellent fit to the data at energies $\lesssim 3$~TeV, where WCDA has sufficient statistics, cf.\ Fig.~\ref{fig:spec-fit}. We therefore adopt the spectrum (\ref{eq:intrinsic}) as the intrinsic one for the fluence $\mathcal{F}$ of the GRB and assume that no new-physics effects modify the absorption in this WCDA energy range.

\vskip 0.5mm
\noindent{\sl 4.3. Results.}
We use 26 data points in the combined fluence spectrum. The standard-physics scenario has no free parameters, while the LIV case introduces one parameter, $E_{\rm LIV, 2}$, and the ALP-mixing case includes two, the axion mass $m$ and the photon coupling $g_{a\gamma\gamma}$. In addition, the ALP model depends on two unknown astrophysical parameters related to the GRB position in its host galaxy: the coordinate along the line of sight, $y_0$, and the orientation angle of the spiral arms, $\phi$ \cite{ST-GRB-hostMF}. For each point $(m, g_{a\gamma\gamma})$ in the ALP parameter space, we average $\chi^2$ over 20 realizations of random $(y_0, \phi)$ values, assuming a uniform distribution of $\phi$ between $0$ and $2\pi$ and the stellar-distribution prior for $y_0$ derived from observations (see Ref.~\cite{ST-GRB-hostMF} for details).

\begin{figure}
    \centering
    \includegraphics[width=\linewidth]{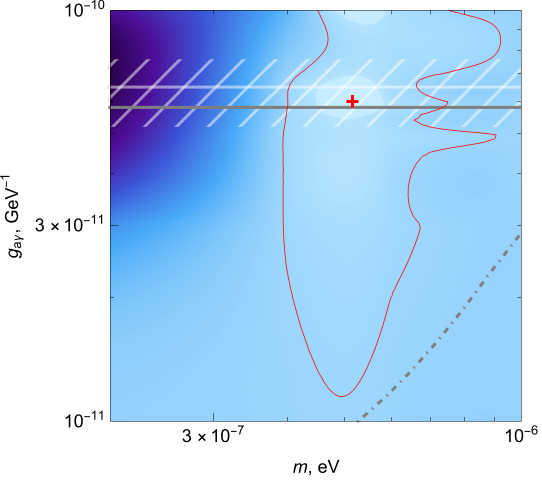}
    \caption{\label{fig:ALPparam} \sl
\textbf{Figure~\ref{fig:ALPparam}.}
ALP parameter space $(m, g_{a\gamma\gamma})$. Blue shading shows the $\Delta\chi^2$ distribution (arbitrary scale; lighter colors indicate better fits). The red cross marks the best-fit point, and the thin red contour outlines the 68\% confidence region. Gray lines indicate existing upper limits on $g_{a\gamma\gamma}$: the solid line is the experimental bound from CAST \cite{CAST:2024}, while the dash-dotted line represents a model-dependent constraint derived from polarization observations of magnetized white dwarfs \cite{MWD-polarization}. The white line with its hatched 68\%~CL band shows the range favored by stellar-evolution arguments \cite{ST-globularCl}.}
\end{figure}

\begin{figure}
    \centering
    \includegraphics[width=0.9\linewidth]{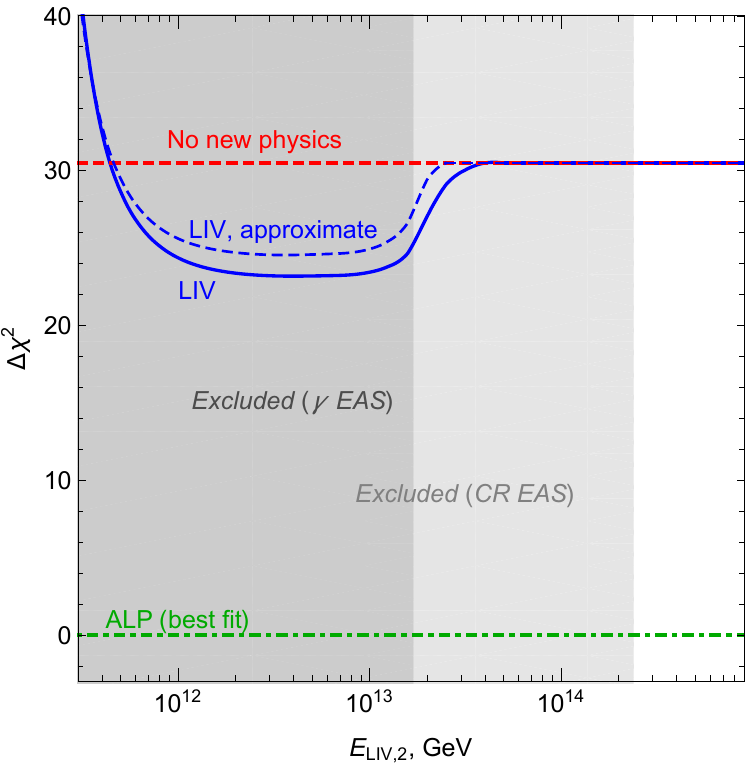}
    \caption{\label{fig:LIVlikely} \sl
\textbf{Figure~\ref{fig:LIVlikely}.}
Comparison of $\Delta\chi^2$ for the quadratic subluminal LIV scenario (solid blue line: exact cross section, Eq.~(\ref{eq:exactCS}); dashed blue line: approximation \cite{Fairbairn:2014LIV}) as a function of $E_{\rm LIV, 2}$. For reference, the horizontal green dash-dotted and red dashed lines indicate the $\Delta\chi^2$ values for the best-fit ALP model, 0,  and the standard-physics case, 30.48, respectively. \red Gray shading indicates the ranges of $E_{\rm LIV, 2}$ excluded from the development of air showers induced by primary gamma rays \cite{Satunin:2021constraints} (dark shade) and cosmic rays \cite{EAS-LIV} (light shade). }
\end{figure}

The results of the fluence-spectrum fits are shown in Fig.~\ref{fig:ALPparam} 
for ALP mixing and in Fig.~\ref{fig:LIVlikely} 
for LIV. The best fit is obtained for the ALP model with $m = 5.16\times10^{-7}$~eV and $g_{a\gamma\gamma} = 6\times10^{-11}$~GeV$^{-1}$, corresponding to an improvement of $\Delta\chi^2 = 30.48$ (24~d.o.f.) over the standard-physics model. For the LIV case, the $\chi^2$ curve exhibits a shallow minimum at $E_{\rm LIV, 2} \sim 4\times10^{12}$~GeV, with a modest improvement of $\Delta\chi^2 = 12.99$ (25~d.o.f.). Model fluence spectra for these best-fit parameters -- using a representative choice of $y_0 = 0$ and $\phi = 80^\circ$ in the ALP case -- are compared with the standard-physics spectrum and the data in Fig.~\ref{fig:spec-fit}.

\vskip 1mm\noindent\textbf{5. Discussion.} 

\begin{figure}
    \centering
    \includegraphics[width=\linewidth]{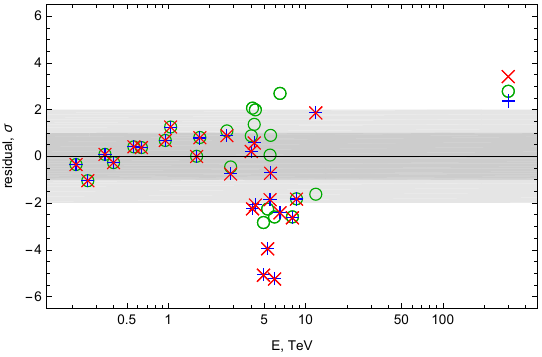}
    \caption{\label{fig:residuals} \red\sl
\textbf{Figure~\ref{fig:residuals}.}
Residuals of fluence spectrum fits for the best-fit ALP (green circles) and LIV (blue pluses) models, and for the standard absorption (red crosses). To combine Gaussian and Poisson residuals in the same plot, we expressed both in terms of ``standard deviations'' $\sigma$, determined in each case from the p-values of individual points. Shading indicates $1\sigma$ and $2\sigma$ bands for reference. For WCDA data points below 2~TeV, new-physics effects in the best-fit models are negligible. The best-fit LIV model differs from the standard physics only for the 300-TeV point.}
\end{figure}

\vskip 0.5mm
\noindent{\sl 5.1. LIV versus ALPs.}
To interpret our results qualitatively, \red consider Fig.~\ref{fig:spec-fit} and Fig.~\ref{fig:residuals}\black. 
The standard absorption model fails to reproduce the Carpet-3 observation and poorly fits the KM2A data if the intrinsic spectrum is assumed to be an unbroken power law. Ref.~\cite{LHAASO-KM2A-GRB} introduced an adjusted absorption model, effectively rescaling the EBL intensity to better match the data. Alternatively, invoking an intrinsic cutoff could explain the $\sim 5$~TeV points, but it would make the $\gtrsim 10$~TeV photons difficult to accommodate.

LIV can only reduce absorption by shifting the pair-production threshold upward and thus cannot reproduce the $\sim 5$~TeV dip in the KM2A spectrum; consequently, the standard and best-fit models coincide below 10~TeV (solid red and dashed black lines in Fig.~\ref{fig:spec-fit}). Photon–ALP mixing, however, introduces oscillations: the best-fit ALP model reproduces the $\sim 5$~TeV dip and enhances the flux near 10~TeV, yielding a much better fit to the KM2A data. At higher energies, the difference between the LIV case (no suppression) and the ALP case ($\sim 1/3$ suppression in the strong-mixing regime) is minor for the contribution of the Carpet-3 point, given its Poisson likelihood with a small but non-zero background (0.003~events). Overall, the ALP scenario provides a clearly superior description of the combined fluence spectrum.

\vskip 0.5mm
\red

\noindent{\sl 5.2. Variations in the assumptions.} To assess the stability of our conclusions with respect to our assumptions, the fit procedure has been repeated for the Franceschini \& Rodighiero (2017) EBL model \cite{Franceschini:2017iwq} and for the magnetic fields of the host galaxy and the Milky Way artificially changed by $\pm 30\%$. The best-fit points for ALP are presented in Fig.~5; they lay well within the $68\%$ CL contour for our baseline model. The best-fit results for LIV do not change, as expected, because it affects only the CMB-absorbed part, not the EBL-absorbed one, and is insensitive to the magnetic field. 
Relaxing the assumption of the power-law spectrum by adding a modest curvature, e.g.\ concave log-parabola, would worsen the fit as explained in Sec.~4.2, but would not change the relative preference of ALPs over LIV. The conclusions of our study remain unchanged with these variations. 
\begin{figure}
    \centering
    \includegraphics[width=\linewidth]{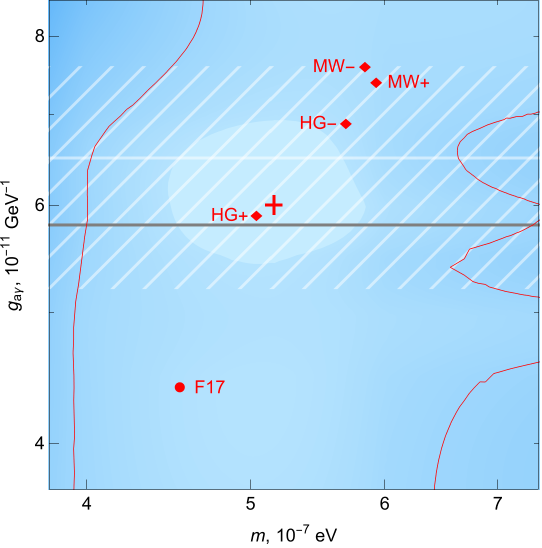}
    \caption{\label{fig:shifts}\red \sl
\textbf{Figure~\ref{fig:shifts}.}
ALP parameter space $(m, g_{a\gamma\gamma})$: zoom in Fig.~\ref{fig:ALPparam}. Best-fit parameters are shown for the baseline model (red cross; red contour encircles the 68\% CL region), variations in the host-galaxy (HG) and Milky-Way (MW) magnetic field normalization by $\pm30\%$ ($+, \ -$), and for the Franceschini \& Rodighiero \ (2017) \cite{Franceschini:2017iwq} EBL model (F17). See Fig.~\ref{fig:ALPparam} for other notations.}
\end{figure}

\black

\vskip 0.5mm
\noindent{\sl 5.3. Other constraints on and indications to LIV.}
Another potential indication of LIV discussed in connection with GRB~221009A concerns the apparent time delay of higher-energy photons relative to lower-energy ones (see, e.g., Ref.~\cite{LHAASO:2024LIVdelay}). In particular, the Carpet-3 event was detected 4536~s after the GRB trigger, significantly later than the $<2000$~s interval covered by the LHAASO detections. Such a delay could originate either from intrinsic emission processes within the source or from subluminal LIV \cite{Ofengeim:2025LIV,Song:2025LIV}. Notably, the LIV mass scale $E_{\rm LIV,2}$ required to account for this time delay is consistent with the value favored by the present spectral analysis.

The same parameter $E_{\rm LIV, 2}$ can be independently constrained by studying air-shower development. In subluminal LIV scenarios, showers initiated by photons above a certain energy are expected to develop deeper in the atmosphere than in standard physics \cite{Rubtsov-et-al-CrossSec}. Observations of very-high-energy gamma-ray spectra of Galactic sources yield a bound of $E_{\rm LIV, 2} > 1.7\times10^{13}$~GeV \cite{Satunin:2021constraints}, while the absence of anomalous photon sub-showers in hadron-induced cascades implies an even stronger limit, $E_{\rm LIV, 2} > 2.4\times10^{14}$~GeV, assuming no LIV for hadrons \cite{EAS-LIV}. These independent constraints disfavor the best-fit $E_{\rm LIV, 2}$ range obtained in our analysis and in Refs.~\cite{Galanti:2022,Finke:2022LIV,Galanti:2025LIVandALP,Ofengeim:2025LIV}.

\red
Note that for $n=1$ LIV, considered in Refs.~\cite{Galanti:2022,Finke:2022LIV,Galanti:2025LIVandALP,Galanti:2025LIV,Song:2025LIV} but not here, the exclusion is actually much stronger. The value of $E_{\rm LIV,1} \sim 3.0\times 10^{20}\,$GeV, preferred by these works, is ruled out if LIV quantum electrodynamics (QED) with $n=1$ is considered self-consistently. The only effective field theory Lagrangian \cite{MyersPospelov} for $n=1$ LIV QED implies  opposite signs for LIV terms in the dispersion relation for two different circular photon polarizations. This implies vacuum birefringence, the rotation of the polarization plane for a photon beam. The strongest constraint related to the absence of vacuum birefringence came from the observation of polarized emission from GRB~061122 by the IBIS telescope of the INTEGRAL mission \cite{Gotz:2013dwa}. It  reads $E_{\rm LIV,1} > 1.8 \times  10^{34}\,$GeV, excluding the value suggested in Refs.~\cite{Galanti:2022,Finke:2022LIV,Galanti:2025LIVandALP,Galanti:2025LIV,Song:2025LIV} for explaining the GRB~221009A by 14 orders of magnitude.
\black

\vskip 0.5mm
\noindent{\sl 5.4. Other constraints on and indications to ALPs.}
The best-fit ALP parameters obtained in this work are consistent with those inferred in previous qualitative analyses of energetic photons from GRB~221009A \cite{ST-GRB-JETPL,ST-GRB-hostMF}, but they clearly favor the high-mass region of the part of the parameter space where strong photon–ALP mixing is expected. The best-fit coupling, $g_{a\gamma\gamma}$, agrees remarkably well with the value favored by the recent reanalysis of evolution of helium-burning stars in globular clusters \cite{ST-globularCl}. It lies close to the experimental upper limit from solar-axion searches with CAST \cite{CAST:2024} and complies with all laboratory constraints. However, it is in tension with the astrophysical bounds derived from polarization measurements of magnetized white dwarfs \cite{MWD-polarization}, which depend sensitively on assumptions about the magnetic-field structure in the vicinity of these stars.

\vskip 1mm\noindent\textbf{6. Conclusions.} 
The combined LHAASO and Carpet-3 observations enable the construction of the fluence spectrum of GRB~221009A over the energy range from $\sim0.1$~TeV to $\sim300$~TeV. We have fitted this spectrum under the assumptions of standard physics, Lorentz-invariance violation, and photon–ALP mixing. Our analysis shows that\red, technically, \black both LIV and ALP scenarios improve the description of the LHAASO and Carpet-3 data compared to standard absorption models\red. However, \black ALP mixing provid\red{}es \black a significantly better overall fit. \red Moreover, the best-fit parameter of LIV, $E_{\rm LIV, 2}$, favored by our analysis, as well as those of Refs.~\cite{Galanti:2022,Finke:2022LIV,Galanti:2025LIVandALP,Ofengeim:2025LIV}, is excluded by other observations. In particular, corresponding values of $E_{\rm LIV, 2}$ would imply that the very air shower detected by Carpet would start deeper in the atmosphere and develop in a non-standard way \cite{Satunin:2021constraints}. We note that $E_{\rm LIV,1}$ values from previous works \cite{Galanti:2022,Finke:2022LIV,Galanti:2025LIVandALP,Galanti:2025LIV,Song:2025LIV} are excluded from non-observation of vacuum birefringence \cite{Gotz:2013dwa}. The ALP best-fit parameters, in turn, \black face tension with model-dependent constraints  \red from magnetic white dwarf polarization \cite{MWD-polarization}\black. These \red tensions \black reinforce the need for new theoretical frameworks to explain the anomalous transparency of the Universe to very high-energy photons. 

\vskip 1mm
\noindent\textbf{Acknowledgements.} This work was supported by the Russian Science Foundation, grant 22-12-00253-P. The authors made use of publicly available large language models to improve the style of some parts of the text of this paper.
\bibliographystyle{nature}
\bibliography{grb}
\end{document}